\documentclass{article}
\usepackage{subequation,graphicx}
\def\circa#1{\,\raise.3ex\hbox{$#1$\kern-.75em\lower1ex\hbox{$\sim$}}\,}
\newcommand{\ov}{{\cal O}}

\newcommand{\f}{{\cal F}}

\newcommand{\be}{\begin{equation}}
\newcommand{\ee}{\end{equation}}
\newcommand{\ben}{\begin{displaymath}}
\newcommand{\een}{\end{displaymath}}
\newcommand{\ba}{\begin{eqnarray}}
\newcommand{\ea}{\end{eqnarray}}
\newcommand{\ban}{\begin{eqnarray*}}
\newcommand{\ean}{\end{eqnarray*}}

\newcommand{\de}{\partial}

\newcommand{\kp}{\mbox{$k$}_{\perp }}
\newcommand{\kt}{\mbox{$k$}_{\perp }}

\catcode`\@=11
 
\begin{document}
\vspace{1cm}

\begin{titlepage}

\vspace*{-64pt}
\begin{flushright}
\end{flushright}

\vskip 1.7cm

\begin{center}
{\Large 
\bf Towards Collinear Evolution Equations\\
in Electroweak Theory
}
\vskip .7cm

{\large Marcello Ciafaloni}

{\it Dipartimento di Fisica, Universit\`a di Firenze
and INFN - Sezione di Firenze,\\via Sansone 1, I-50019 Sesto Fiorentino,
Firenze (Italy),\\
E-mail: ciafaloni@fi.infn.it}

\vskip.5cm

{\large Paolo Ciafaloni}

{\it INFN - Sezione di Lecce,
\\Via per Arnesano, I-73100 Lecce, Italy
\\ E-mail: Paolo.Ciafaloni@le.infn.it}

\vskip.5cm

{\large Denis Comelli}

{\it INFN - Sezione di Ferrara,
\\Via Paradiso 12, I-35131 Ferrara, Italy\\
E-mail: comelli@fe.infn.it}
\end{center}

\vskip .5cm

\begin{abstract}

We consider electroweak radiative corrections to hard inclusive processes
at the TeV scale, and we investigate how collinear logarithms factorize in
a spontaneously broken gauge theory, similarly to the DGLAP
analysis in QCD.
 Due to the uncancelled double logs
noticed previously, we find a factorization pattern which is qualitatively
different from the analogous one in QCD.
New types of splitting functions
emerge which are needed to describe the initial beam charges and are
infrared-sensitive, that is dependent on an infrared cutoff provided,
ultimately, by the symmetry breaking scale. We derive such splitting
functions at one-loop level in the
example of $SU(2)$ gauge theory, and we also discuss the structure 
functions' evolution equations, under the assumption that 
isospin breaking terms present in the Ward identities of the theory 
are sufficiently subleading at higher orders.    

\end{abstract}

\end{titlepage}

%

\section{Introduction}

The recent analysis of enhanced electroweak corrections at the TeV
scale \cite{cc} 
has shown that, for $ s \gg M^2$, even inclusive observables take 
large contributions from both collinear and infrared singularities,
arising in the limit of vanishing symmetry breaking scale \cite{ccc}.
The magnitude of such enhancements, involving the effective coupling
$\alpha_w\log^2\frac{s}{M_W^2}$, 
raises the question of taking into account single logarithmic collinear
contributions, a problem which corresponds to the usual DGLAP \cite{DGLAP}
analysis in QCD. 

In a spontaneously broken gauge theory, the analysis of mass singularities  
is complicated by two peculiar features: firstly, the fact that initial states,
like electrons and protons in the accelerator beams, carry nonabelian 
(isospin) charges and may be mixed charge states \cite{abelian}. This feature,
due to symmetry breaking at low
energies,  causes the very existence of
double logs, i.e. the lack of cancellation of virtual corrections with real
emission in inclusive observables. Secondly, the gauge theory Ward Identities
are broken by Goldstone boson contributions \cite{WI} which are, however, 
proportional to
the symmetry breaking scale and are therefore expected to be
suppressed, to a certain extent, in the high energy limit. 
In a series of papers \cite{ccc} we have analyzed the problem above at the 
leading double-log level by using, for $\sqrt{s}\gg M_W$, Ward
Identity restoration, and therefore isospin and charge conservation in the  
squared matrix elements. At this level, double-log effects are classified
according to the total isospin $T$ and/or total charge in the $t$-channel of 
the inclusive overlap matrix being considered.       

In this paper we analyze, under the same assumptions, collinear singularities
at leading single-logarithmic level 
and we argue that Ward identities are restored at least at one-loop level:
this is sufficient in order to 
provide the analogue of DGLAP splitting functions for a broken gauge theory. 
A key feature of the result is the existence of a new kind of splittings,
which correspond to a non-vanishing t-channel isospin $T\neq 0$ and are 
infrared singular, in the sense that they are explicitly dependent on the
infrared cutoff provided by the symmetry breaking scale, and are thus 
responsible for the double logs. We
derive here the set of splitting functions in the example of a
 spontaneously broken  SU(2) gauge theory. We limit ourselves to the case of
initial light fermions and transverse bosons; longitudinal bosons and the Higgs
sector will be considered elsewhere \cite{toappear}. By further assuming, in
this example, collinear
factorization to all orders we also discuss the solution of the ensuing 
evolution equations, including the novel ones with $T\neq 0$. 
Ultraviolet logarithms leading to running coupling effects are neglected
throughout the paper.

\section{Collinear Ward Identities}

Consider for definiteness a lepton initiated Drell Yan process
of type $e^+(p_1)\;e(p_2)\rightarrow q(k_1) \bar{q}(k_2)  +X$ where 
$s=2 p_1 \cdot p_2$ is the total invariant mass and $Q^2=2 k_1 \cdot k_2$
 is the hard scale.
The differential cross section of such process is related to the overlap
function\footnote{Here and in the following 
we adopt definitions  and conventions
used in \cite{ccc}.}
${\cal O}_{\alpha_1 \beta_1, \alpha_2 \beta_2}(p_1,p_2,k_1,k_2)$
for $\alpha_1=\beta_1=e$  and $\alpha_2=\beta_2=e^+$, and
we are interested in the collinear singular radiative corrections 
to the Born level overlap function ${\cal O}^H$ .

\begin{figure}
      \centering
      \includegraphics[height=80mm]
                  {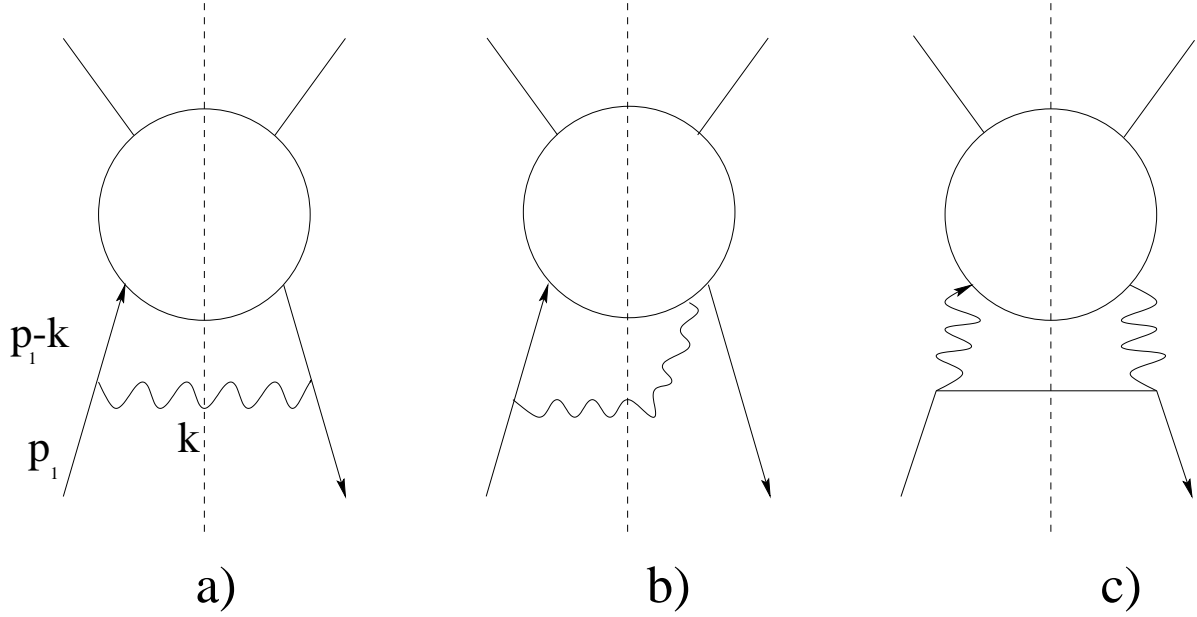}
      \caption{\label{fig1} Unitarity diagrams for one-loop real emission 
corrections to the inclusive hard overlap matrix $\ov^{Hf}$. The symmetrical
      counterpart of diagram $b)$ is not shown. }
      \end{figure}

In order to compute such corrections, we first concentrate on the behavior 
of the contribution of an emitted  boson 
closely parallel to an initial fermion leg $p_1$. 
In the Feynman 't Hooft gauge three kinds of unitarity  diagrams
contribute to collinear singularities (Fig. \ref{fig1}). 
The contributions in Fig. \ref{fig1}(a), \ref{fig1}(c) 
contain a double pole factor due to the denominators $2 p_1\cdot k$
and can be explicitly evaluated by performing the usual Dirac algebra.
The one in Fig \ref{fig1}(b), with a single denominator,  
includes the sum of vector boson insertion diagrams on all interfering legs 
but $p_1$. It
can be evaluated in the collinear limit $k_{\mu}\,\propto\,p_{1 \mu}$
thanks to the gauge theory Ward Identities.

\begin{figure}
      \centering
   \includegraphics[width=15cm]
                  {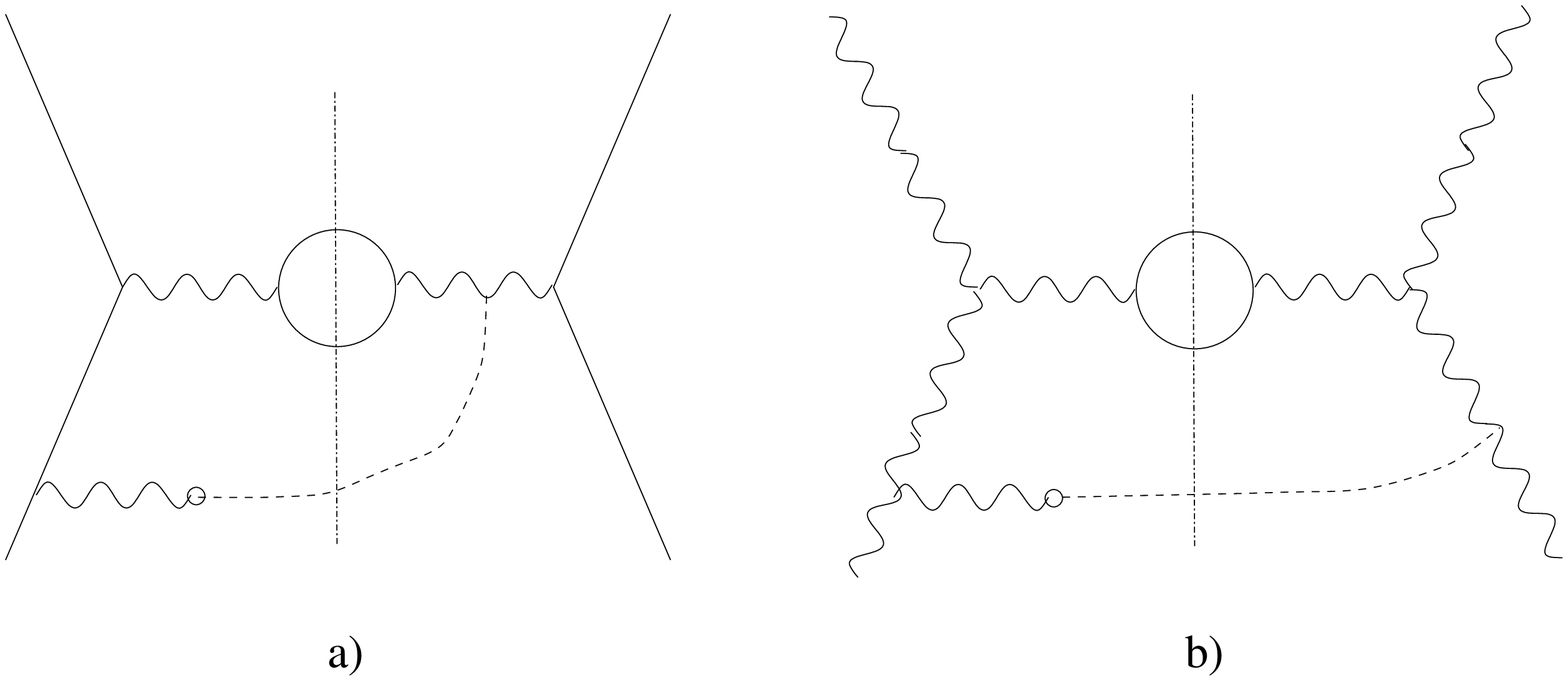}
      \caption{\label{fig3}  Examples of one-loop Goldstone boson
      contributions to CWI for (a) fermion-initiated and (b) boson-initiated
      processes. The wavy (dashed) lines label vector boson (Goldstone boson)
       exchanges, the small circles denote the $M$ couplings arising from the
      use of CWI's.}
      \end{figure}

The method of the collinear Ward Identities (CWI), introduced by
Amati, Petronzio and Veneziano in QCD \cite{acv} consists in factorizing the
contributions in Fig \ref{fig1}(b) 
by relating them to the remaining insertion on the outgoing
$p_1$ leg, plus Goldstone boson contributions. At one-loop level, the latter
are of the type in Fig. \ref{fig3}, carrying a coupling of order M (the
symmetry breaking scale), and are suppressed 
by a power $M^2/s$ for initial 
light fermions, which do not couple explicitly to Goldstones (Fig. 2(a)).
On the other hand, for initial transverse  bosons, the suppression factor is
rather $M^2/k_T^2$ (Fig. \ref{fig3}(b)); 
therefore, upon $k_T$ integration, they do not
provide any collinear logs at all. Goldstone contributions are thus negligible
in either case, at this level.

 By then using unbroken CWI, the diagrams in Fig. 1(a,b), coupling the
fermionic overlap matrix with itself, are evaluated as follows
\be\label{ciccia}
\delta\ov^f_{\alpha_1\beta_1}=
\frac{\alpha_W}{2\pi}
\int_{M^2}^s\frac{d\kt^2}{\kt^2}
\int_0^{1-\frac{\kt}{\sqrt{s}}}
\frac{dz}{z}\left[(1-z)+\frac{2z}{1-z}\right]
\left[t^a\ov^{Hf}(zp_1)t^a\right]_{\alpha_1\beta_1}
\ee
 where $t^a$ denote the isospin matrices in the fundamental representation,
and the first (second) term in square brackets correspond to 
Fig. \ref{fig1}(a)
(\ref{fig1}(b)) respectively. 
Let us note that the longitudinal fraction $1-z$ is here cutoff by
$\epsilon = \frac{\kt}{\sqrt{s}}$. The precise value of such cutoff is 
important for
disentangling double and single logarithms and can be derived as follows.
The emitted boson admits the Sudakov parameterization
$k^\mu=(1-z)p_1^\mu+\bar{z}p_2^\mu+\kt^\mu;\;
\bar{z}=\frac{\kt^2+M^2}{(1-z)s}$. It is clear then that the angular 
region
with the boson emitted in the forward hemisphere defined by $p_1$ 
corresponds to $1-z>\bar{z}=\frac{\kt^2+M^2}{(1-z)s}$, so that we can set
$1-z>\frac{\kt}{\sqrt{s}}>\frac{M}{\sqrt{s}}$, which are the integration
boundaries appearing in (\ref{ciccia}).
It is important to note that 
eq.  (\ref{ciccia}) was derived in the collinear region
$1-z\gg\frac{\kt}{\sqrt{s}}$, but is still valid in the infrared region
$1-z=O(\frac{\kt}{\sqrt{s}})$ because of the eikonal approximation derived in
our previous work. Therefore the precise value of the cutoff is well
determined by the above argument.

By including the contributions of fig. \ref{fig1}(c) 
 and
virtual contributions \cite{Denner}, we obtain
\begin{equation}\label{cicciotto}
           \begin{eqalign}   
\delta\ov^f_{\alpha_1\beta_1}&=
\frac{\alpha_W}{2\pi}\int_{M^2}^s\frac{d\kt^2}{\kt^2}
\int_0^{1}
\frac{dz}{z}\left\{
P^R_{ff}(z)\theta(1-z-\frac{\kt}{\sqrt{s}})
\left[t^a\ov^{Hf}(zp_1)t^a\right]_{\alpha_1\beta_1}
\right.\\
&+\left.
P^R_{bf}(z)\left[t^A\ov^{Hb}_{AB}(zp_1)t^B \right]_{\alpha_1\beta_1}
+ C_f 
P^V_{ff}(z,\frac{\kt}{\sqrt{s}}) \left[\ov^{Hf}(z p_1)\right]_{\alpha_1\beta_1}
\right\}
\end{eqalign}\ee
\be \label{pff}
P^R_{ff}(z)=\frac{1+z^2}{1-z}
\qquad
P^R_{bf}(z)=\frac{1+(1-z)^2}{z}
\qquad
P^V_{ff}(z,\epsilon)=
-\delta(1-z)(\log\frac{1}{\epsilon^2}-\frac{3}{2})
\ee
The real-emission splitting function $P_R(z)$ occurring in 
eq. (\ref{cicciotto})
is the same occurring in the unbroken theory; this is not surprising since we
have assumed restored collinear Ward Identities. On the other hand the broken
theory differs for the presence of the cutoff $1-z>\frac{\kt}{\sqrt{s}}$ and
because of the isospin structure. In fact there is no isospin averaging on the
initial state, in contrast to the corresponding (unbroken) QCD case where 
physical quantities are averaged over colour.

The one loop formula (\ref{cicciotto}) is consistent
 with a general factorization formula
of type
\be\label{ff}
\ov_{ij}(p_1,p_2;k_1,k_2)=\int \frac{dz_1}{z_1}
\frac{dz_2}{z_2}\f_{ki}(z_1;s,M^2)
\ov^H_{k l}(z_1 p_1,z_2 p_2;k_1,k_2) \f_{lj}(z_2;s,M^2)
\ee
where $i,j$ label the  kind of particle (fermion, gauge  boson),
 and  where
isospin indices in the overlap function $\ov$ and structure function $\f$ are 
understood.
Of course we have not proved eq. (\ref{ff}) 
because this requires neglecting Goldstone bosons insertions at higher orders 
or better proving that they are subleading by at least two logarithms
 \cite{toappear}.

\section{Evolution Equations}

If nevertheless the factorization formula (\ref{ff}) is assumed, and 
CWI are supposed to be restored at higher orders as well,
the structure functions will satisfy evolution equations with respect to 
a collinear  cutoff
$\mu$ parameterizing the lowest value of $\kt$, as follows
(see fig. \ref{fig2}; $t=\log\mu^2$):
\begin{subequations}\label{kpil}\begin{eqalignno}
-\frac{\de \f_{if}^{\alpha\beta}}{\de t}&=
\frac{\alpha_W}{2\pi}\left\{
C_f \f_{if}^{\alpha\beta} \otimes P^V_{ff}
+\left(t^C\f_{if}t^C\right)^{\alpha\beta}\otimes P_{ff}^R
+\left(t^A\f_{ib}^{AB}t^B\right)_{\alpha\beta}\otimes P^R_{bf}
\right\}
\\
-\frac{\de \f_{ib}^{AB}}{\de t}&=
\frac{\alpha_W}{2\pi}
\left\{C_b  \f_{ib}^{AB}\otimes P^V_{bb}+
\left(T^C\f_{ib}T^C\right)^{AB}\otimes P_{bb}^R
+{\rm Tr}(t^A\f_{if}t^B)\otimes P^R_{fb}
\right\}
\end{eqalignno}\end{subequations}
In these equations $T^a$ ($t^a$) denote the isospin matrices
 in the adjoint (fundamental)
representation and $\f_{ij}^{\alpha\beta}$ denotes the distribution of a
particle $i$ (whose isospin indices are omitted)
inside particle $j$ (with isospin
leg indices ${\alpha,\beta}$); the trace is taken,
here and in the following, with respect to the indices of the soft scale leg
$j$. Since the index $i$ is always kept fixed, we will omit it from now on,
with the understanding that, for istance, $\f_f$ collectively
denotes 
all $\f_{if}$ with any value of $i$. Furthermore, we have defined
the convolution
$
[f\otimes P](x)\equiv\int_x^1 P(z)f(\frac{x}{z})\frac{dz}{z}
$. The relevant distributions are 
 $P^{V}_{ff}(z,\frac{\mu}{\sqrt{s}}),
P^{R}_{ff}(z), {P}^R_{bf}(z)$ given in eq.(\ref{pff})
and $P_{bb}^R(z),P_{fb}^R(z),P_{bb}^V(z,\frac{\mu}{\sqrt{s}})$ where
\be
P_{bb}^R(z)=
2\left(
z(1-z)+\frac{1-z}{z}+\frac{z}{1-z}
\right);
\;\;\;\;
P_{fb}^R(z)=z^2+(1-z)^2;
\;\;\;\;
P_{bb}^V(z,\epsilon)=-\delta(1-z)\left(
\log \frac{1}{\epsilon^2}-\frac{11}{6}+\frac{n_f}{6}
\right)
\ee
Here $n_f$ is the number of fermion doublets, and the real emission cutoff
 $1-z>\epsilon=\mu/\sqrt{s}$ is understood.

\begin{figure}
     \centering
     \includegraphics[height=80mm]
                  {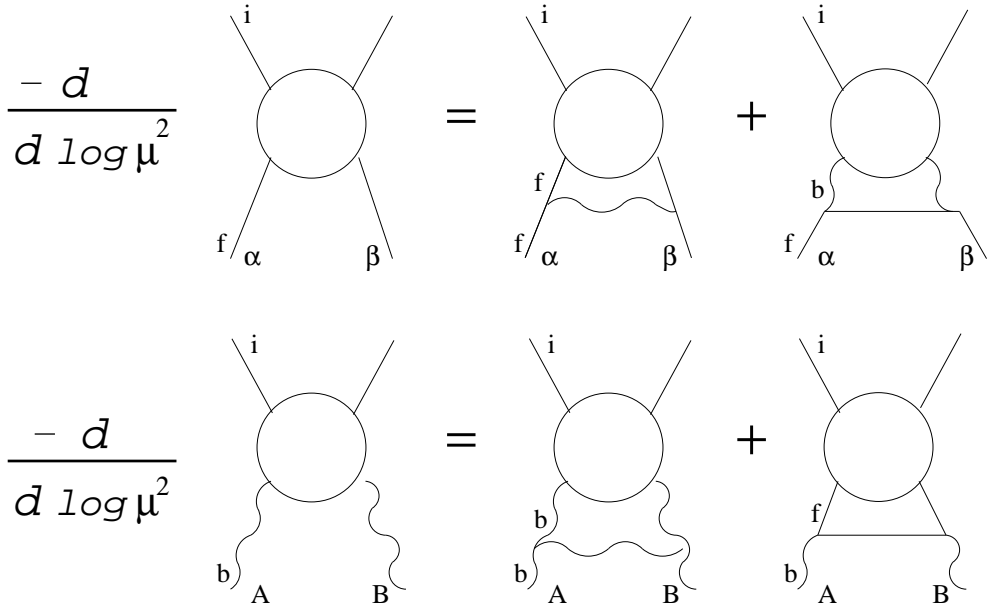}
      \caption{\label{fig2} Picture of the collinear evolution equations of
     eq. (\ref{sys}) in the light-fermion, vector-boson sector. }
      \end{figure}

Because of the assumed recovered isospin invariance, eq. (\ref{ff}) 
can be decomposed in $t$-channel isospin  components
as follows:
\be\label{overlap}
\ov_{ij}(p_1,p_2;k_1,k_2)=\int \frac{dz_1}{z_1}
\frac{dz_2}{z_2}\sum_{T}^{0,1,2,...}\,f^{(T)}_{ki}(z_1;s,M^2)\,     
\ov^{H(T)}_{k l}(z_1 p_1,z_2 p_2;k_1,k_2) \,f^{(T)}_{lj}(z_2;s,M^2)
\ee
Then, by using the following  projections, with a definite value of the
t-channel isospin $T=0,1,2$ \cite{ccc}
\be\label{tt}
f^0_f=\frac{1}{2}{\rm Tr}\{\f_{f}\}=\frac{f_e+f_\nu}{2}\quad
f^1_f={\rm Tr}\{\f_{f}t_3\}=\frac{f_\nu-f_e}{2}\quad
f^0_{\bar{f}}=\frac{f_{\bar{e}}+f_{\bar{\nu}} }{2}\quad
f^1_{\bar{f}}=\frac{f_{\bar{e}}-f_{\bar{\nu}}}{2} 
\ee\be
f^0_{b}=\frac{f_++f_3+f_-}{3}\quad
f^1_{b}=\frac{f_+-f_-}{2}\quad
f^2_{b}=\frac{f_++f_--2f_3}{6}
\ee
the operatorial equation (\ref{kpil}) 
can be decomposed as a system of scalar equations 
with definite $T$ values:
\begin{subequations}\label{sys}\begin{eqalignno}  
-\frac{\de f_{f}^0}{\de t}&=\frac{\alpha_W}{2\pi}\left\{
\frac{3}{4}f_{f}^0\otimes(P_{ff}^R+P^V_{ff})+\frac{3}{4}f^0_{b}\otimes
P^R_{bf}\right\}
\\
-\frac{\de f_{b}^0}{\de t}&=\frac{\alpha_W}{2\pi}\left\{
2f_{b}^0\otimes(P_{bb}^R+P^V_{bb})+\frac{1}{2}
(f^0_{f}+f^0_{\bar{f}})\otimes
P^R_{fb}\right\}
\\
-\frac{\de f_{f}^1}{\de t}&=\frac{\alpha_W}{2\pi}\left\{
f_{f}^1\otimes P^V_{ff}-
\frac{1}{4}f_{f}^1\otimes(P_{ff}^R+P^V_{ff})+\frac{1}{2}f^1_{b}\otimes
P^R_{bf}\right\}
\\
-\frac{\de f_{b}^1}{\de t}&=\frac{\alpha_W}{2\pi}\left\{
f_{b}^1\otimes P^V_{bb}+
f_{b}^1\otimes(P_{bb}^R+P^V_{bb})+\frac{1}{2}(f^1_{f}+f^1_{\bar{f}})
\otimes
P^R_{fb}\right\}
\\
-\frac{\de f_{b}^2}{\de t}&=\frac{\alpha_W}{2\pi}\left\{
3f_{b}^2\otimes P^V_{bb}-
f_{b}^2\otimes(P_{bb}^R+P^V_{bb})\right\}
\end{eqalignno}\end{subequations}
with similar equations holding for $f^T_{\bar{f}}$.
Notice that for $T=0$, real and virtual contributions carry the same charge
factor so that the distribution $ P^R_{ii}(z)+P^V_{ii}(z)$
is regularized and the cutoff
$\frac{\kt}{\sqrt{s}}$ is irrelevant in the strong-ordering region $M\ll
\kt\ll \sqrt{s}$.  On the other hand for $T\neq 0$ real and
virtual emission do not cancel and the double-log behavior due to the
singularity at $z=1$ is exposed. This lack of compensation, due to 
the fact that no averaging over isospin quantum numbers is done \cite{ccc}, is
apparent in the first (virtual) term of eqs. (\ref{sys}), for $T$ = 1,2.  

Eqs. (\ref{sys}) can be partly diagonalized by introducing the valence--like
quantities $V^T=\frac{f^T_f-f^T_{\bar{f}}}{2}$, in terms of which we have:
\be
-\frac{dV^0}{dt}=\frac{\alpha_W}{2\pi}\,\frac{3}{4}V^0\otimes 
(P^R_{ff}+P^V_{ff})\qquad
-\frac{dV^1}{dt}=\frac{\alpha_W}{2\pi}\,V^1\otimes\left\{
 P_{ff}^V-\frac{1}{4}\,(P^R_{ff}+P^V_{ff}) \right\}
\ee
Therefore, $f^2_b$,$ V^0$ and $V^1$ satisfy single-channel evolution
equations, while
the remaining ones couple sea--like and boson distributions.
Notice however that, for $T = 1$ these are not the usual ``valence'' or ``sea''
combinations appearing in QCD, because of the $t_3$ weight in eq. (\ref{tt}). 
For instance we have:
\be
V^0=\frac{f_\nu-f_{\bar{\nu}}}{2}+\frac{f_e-f_{\bar{e}}}{2}\qquad
V^1=\frac{f_\nu+f_{\bar{\nu}}}{2}-\frac{f_e+f_{\bar{e}}}{2}
\ee

The explicit solution of eqs. (\ref{sys}) is obtained as 
in QCD for $T=0$ where only single logarithms occur.
For $T\neq 0$, on the other hand, it is always possible to extract the 
double-log contribution by defining new
distributions $\tilde{f}$ whose evolution equations are regular 
and can be solved as in QCD. For instance, for fermions we set:
 \begin{equation}\label{tilde}
           \begin{eqalign}
f_{f}^T(x;s,M^2)&=
\exp \left[\frac{\alpha_W}{4\pi}\frac{T(T+1)}{2}
\int_0^1dz\int_{M^2}^s\frac{d\kp^2}{\kp^2}{P}^V_{ff}(z,\frac{\kt}{\sqrt{s}})
\right] \tilde{f}_{f}^T(x;s,M^2)
\\
&=
\exp\left[-\frac{\alpha_W}{4\pi}\frac{T(T+1)}{2} (\log^2 \frac{s}{M^2}-3
\log \frac{s}{M^2})\right] \tilde{f}_{f}^T(x;s,M^2)
           \end{eqalign}
     \end{equation}
where we have taken $\kt=M$ ($\kt=\sqrt{s}$) as lower (upper) bound of the
$\kt$ integration.\footnote{One could have taken $\kt=Q$ instead of
 $\kt=\sqrt{s}$ as upper bound. This difference
does not contribute any logarithms to eq. (\ref{tilde}), because
$Q/\sqrt{s}=O(1)$, and is therefore a subleading effect.}It is then 
straightforward
to check that indeed the $\tilde{f}_{f}^T$'s satisfy regularized evolution
equations, whose splitting functions are read off in the second term of eqs.
(\ref{sys}) for $T=1$.

Finally, the overlap matrix $\ov_{ij}$ itself is determined from
eq. (\ref{overlap}), on the basis of the hard overlap matrix
$\ov_{kl}^{H(T)}$, to be
calculated perturbatively for the various $T$ values, given the initial
isospin states of the process.

To summarize, we have derived the splitting functions and written down the 
evolution equations for radiative
corrections of electroweak type in a spontaneously broken theory. Our 
analysis in the various 
t-channel isospin $T$ components shows that for $T$=0 the equations are quite
similar to the DGLAP ones in QCD. However, for $T\neq 0$, the infrared 
singularity is exposed, originating double logarithms that are
absent in the unbroken theory.
Further developments require showing that indeed Goldstone boson insertions
in the Ward Identities can be neglected at higher orders, 
as assumed in this paper. They also require considering 
longitudinal degrees of freedom, which mix with Higgs states and have a more
complicated group structure, as noticed in \cite{ccc1} at double-log level. 
But, on the whole, the route for writing the collinear evolution equations in 
the full Standard Model is open.

\end{document}